\documentclass{elsart}

 \usepackage{graphicx,psfrag,multirow}

\usepackage{amssymb}
\usepackage{amsmath}
\usepackage{amsmath,amsfonts,amssymb}

\begin{document}

\begin{frontmatter}




\title{Physics inspired algorithms for (co)homology computation}

\author[Krakow]{Pawe{\l}~D{\l}otko\thanksref{pawel}},
\thanks[pawel]{Partially supported by grant IP 2010 046370.}
\ead{pawel.dlotko@uj.edu.pl}
\author[Udine]{Ruben Specogna\corauthref{cor}},
\corauth[cor]{Corresponding author. Phone: +39 0432 558037. Fax: +39 0432 558251.}
\ead{ruben.specogna@uniud.it}
\address[Krakow]{Jagiellonian University, Institute of Computer Science, Lojasiewicza 4, 30348 Krak\'{o}w, Poland}
\address[Udine]{Universit\`a di Udine, Dipartimento di Ingegneria Elettrica, Gestionale e Meccanica, Via delle Scienze 208, 33100 Udine, Italy}

\hyphenation{sem-i-con-duct-ors}

\begin{abstract}
The issue of computing (co)homology generators of a cell complex is gaining a pivotal role in various branches of science.
While this issue can be rigorously solved in polynomial time, it is still overly demanding for large scale problems.
Drawing inspiration from low-frequency electrodynamics, this paper presents a physics inspired algorithm for first cohomology group computations on three-dimensional complexes. The algorithm is general and exhibits orders of magnitude speed up with respect to competing ones, allowing to handle problems not addressable before.
In particular, when generators are employed in the physical modeling of magneto-quasistatic problems, this algorithm solves one of the most long-lasting problems
in low-frequency computational electromagnetics. In this case, the effectiveness of the algorithm and its ease of implementation may be even improved by introducing the novel concept of \textit{lazy cohomology generators}.
\end{abstract}

\begin{keyword}
physics inspired algorithms \sep algebraic topology \sep (co)homology \sep computational physics \sep magneto-quasistatics \sep eddy-currents
\PACS 41.20.Gz \sep 02.40.Re \sep 02.70.Dh \sep 07.05.Tp
\end{keyword}
\end{frontmatter}

\section{Introduction}
The availability of unprecedented computing power and efficient numerical methods produced a dramatic increase in applications of computational (co)homology~\cite{munkres} (i.e. the computation of generators of the (co)homology group.) (Co)homology, in fact, has been already shown to be essential in unexpected areas of science, ranging from computer aided design (CAD) for feature detection~\cite{dey1}, 
parametrization and mesh generation~\cite{gu3}, shape analysis and pattern
recognition~\cite{pattern}, to sensors networks~\cite{sensors} and robot
motion planning~\cite{robotmotionplanning}, medicine and biology~\cite
{cancer} and quantum chemistry~\cite{mezey}. Focusing on
physics, (co)homology generators have been used for example to detect the
chaotic behavior in sampled experimental data~\cite{k1},~\cite{k2},~\cite
{k3}, in quantum information theory~\cite{distillation}, string theory~\cite
{jurke} and electromagnetism~\cite{kotiuga},~\cite{sst},~\cite{cmame},~\cite
{cmes},~\cite{ijnme},~\cite{CiCP}.

The problem addressed in this paper is the computation of the first cohomology group generators for three-dimensional combinatorial manifolds, being the computation of the zeroth and second (co)homology groups already satisfactorily solved in literature.

Cohomology generators over integers---unlike the real and complex ones---can be rigorously computed in polynomial time by finding the \textit{Smith normal form} (SNF)~\cite{munkres} of the coboundary matrix. However, this approach is computationally not attractive because the best implementation of the SNF exhibits a hyper-cubical complexity~\cite{bestSmith}. Sparse matrix data structures and the \textit{reduction} of the input complex~\cite{kotiuga},~\cite{cmes} before the SNF computation is run have been used to speed up the computation to a point that can be used for practical problems.
Recently, a novel algorithm called \textit{Thinned Current Technique} (\texttt{TCT}) has been introduced in~\cite{tct}. This algorithm exhibits orders of magnitude speed up on typical problems with respect to other algorithms presented in literature and it is easy to implement. Nonetheless, its main limitation is that when the complex is not skeletonizable---i.e. cannot be homotopically retracted to a graph---this algorithm uses again the technique presented in~\cite{cmes}. For a comprehensive survey on other algorithms proposed in literature refer to~\cite{CiCP}.

To introduce a physics inspired algorithm for first cohomology group computations, let us focus on the interplay between (co)homology, discrete Hodge decomposition~\cite{bott},~\cite{kotiuga} and physical modeling that appears when considering problems whose definition of potentials is not straightforward. One of the most studied examples where this happens occurs in low-frequency electrodynamics. Electromagnetic phenomena are governed by Maxwell's laws~\cite{maxwell} and material constitutive relations. For slowly time-varying fields, whose change in magnetic field energy is dominant and electromagnetic wave propagation can be ignored, it is typical to brake the symmetry of Maxwell's laws by neglecting the displacement current in the Amp\`{e}re--Maxwell's equation~\cite{maxwell}. Using this magneto-quasistatic (MQS) approximation, the magnetic field is irrotational in the insulators thanks to Amp\`{e}re's law. The fact that the insulating region is in most cases not simply connected prevent it to be exact. The consequence is that a magnetic scalar potential cannot be introduced na\"{\i}vely in the insulating regions. Yet, using a scalar potential is 
tempting since formulations based on it are computationally much more efficient than the ones using the classical magnetic vector potential.

How to define a magnetic scalar potential in non simply connected domains has drawn a considerable effort in the computational electromagnetics community in the last twenty-five years. A connection of this issue with (co)homology theory has been advocated many years ago~\cite{kotiugaap},~\cite{kotiuga}.
Despite most scientists keep using other heuristic and sometimes patently incorrect definition of potentials and related algorithms (see references and counter-examples in~\cite{CiCP}), it has been recently shown that the first cohomology group generators over integers of the cell complex modeling the insulating region are needed to make the problem well defined~\cite{cmame},~\cite{CiCP}. Introductory material on this subject comprising an informal introduction on algebraic topology, how to model physical variables as cochains with complex coefficients and how to rephrase Maxwell's laws in algebraic form, can be found in~\cite{cmame},~\cite{CiCP},~\cite{tct}.

The quest for an algorithm for first cohomology group computation that is both general and exhibits a linear average complexity is still open. This is surprising since the research on this issue has been pushed forward by many leading software houses having at least part of the core business in solving MQS problems as MagSoft Corp., Ansoft Corp.--ANSYS Inc., Vector Fields L.t.d., CST Ag., and Comsol Inc.. The fact that this issue has been considered unsolved for so many years indicates that computing cohomology generators quickly (and correctly) is not straightforward. Also the implementation complexity may affect negatively the technology transfer.
Developing a simple and fast algorithm would enable to embed it in the next-generation of electromagnetic Computer-Aided Engineering (CAE) softwares.

This article fills this gap by exploiting a novel physics inspired approach to compute cohomology generators suitable for physical modeling called the \textit{D{\l}otko--Specogna} (\texttt{DS}) algorithm. Moreover, the novel concept of \textit{lazy cohomology generators} is introduced to speed up
and simplify the implementation when cohomology generators are employed in computational physics.

The paper is structured as follows. Section~\ref{sec:introToAT} is a mild introduction to (co)homology theory that may be skipped by readers familiar with this topic. In Section~\ref{cohomologyinem} the role of (co)homology theory in low-frequency electrodynamics is recalled.
The \textit{D{\l}otko--Specogna} (\texttt{DS}) algorithm and the novel concept of \textit{lazy cohomology generators} are introduced in Section~\ref{sec:DSAlgorithm}. In Section~\ref{numres} some numerical experiments are presented to compare the novel algorithm to other state-of-the-art algorithms in terms of efficiency and robustness. Finally, in Section~\ref{conclusions}, the conclusions are drawn.

\section{Mild introduction to algebraic topology}
\label{sec:introToAT}
In this section a mild introduction to algebraic topology is provided. For a rigorous one please consult~\cite{massey}. Let us consider a discretization $\mathcal{K}$ of a given three-dimensional space as cell complex (more precisely, a regular CW-complex~\cite{massey}.) For simplicity, one may think about a simplicial complex.
$\mathcal{K}$ is a \emph{combinatorial manifold} if the link of every vertex is a sphere or disk. The link of a vertex $v \in \mathcal{K}$ consists of all elements $s \in \mathcal{K}$ that do not contain $v$ that are faces of higher dimensional elements containing $v$.

Homology and its dual cohomology theory are mathematical tools to describe ``holes'' of a given space in a rigorous way.
\begin{figure}
\centering
\includegraphics[width=10cm]{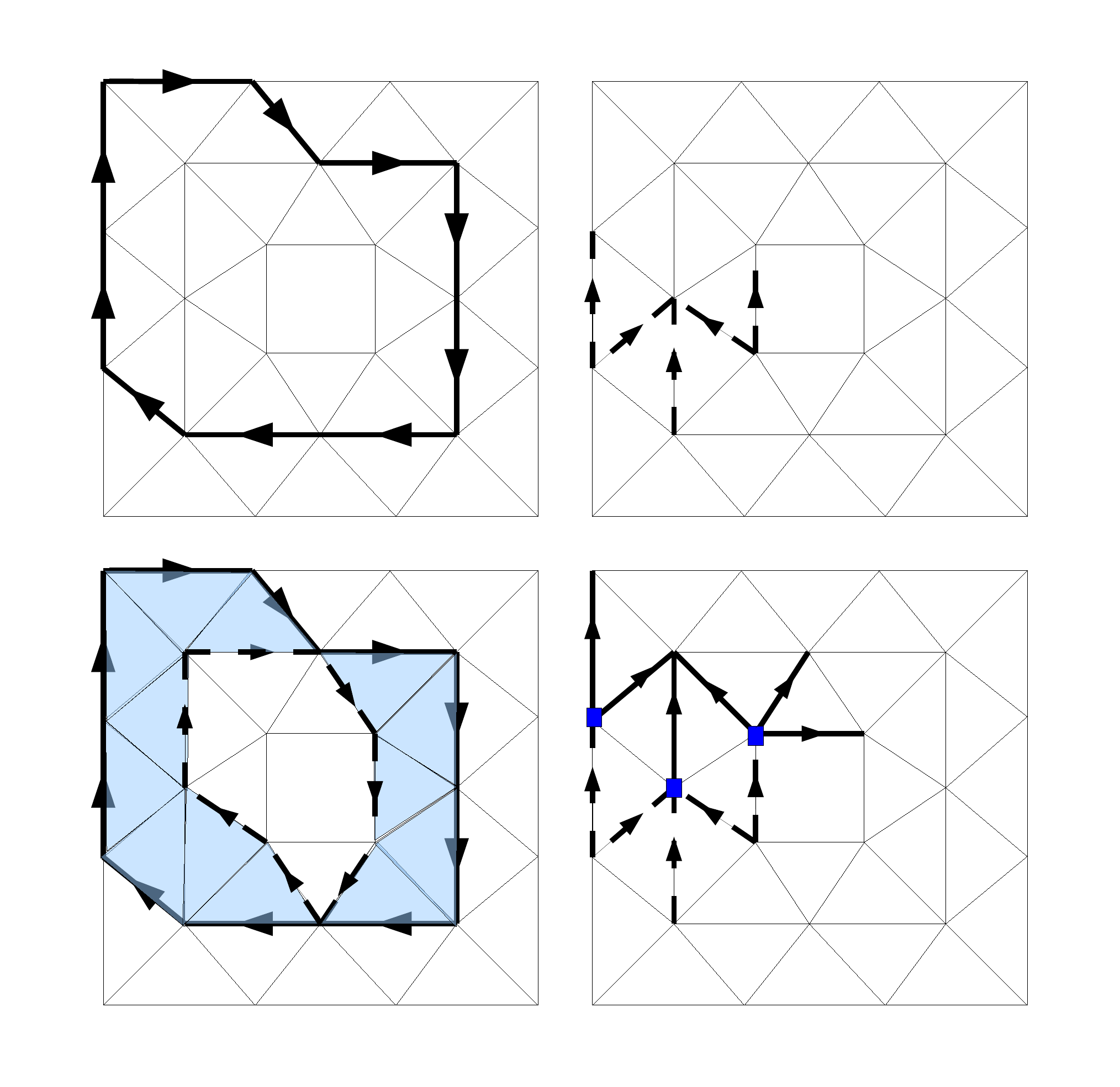}
\caption{On the left (upper and lower), representants of homology generators of the annulus. On the right (upper and lower), representants of cohomology generators of the annulus.}
\label{fig:hom}
\end{figure}

As an example, let us consider the two-dimensional simplicial complex $\mathcal{K}$ of the annulus represented in the Fig.~\ref{fig:hom}. The zero-dimensional holes are defined as the connected components of $\mathcal{K}$. In the example, $\mathcal{K}$ is formed by a single connected component.
Clearly, there is one (one-dimensional) hole in the annulus. This hole can be surrounded with a one-dimensional oriented cycle, as in Fig.~\ref{fig:hom} upper left. This kind of cycle represents a generator of the first homology group of the annulus.

The concept of holes may be generalized to higher dimensions. In the example, there are no holes of dimensions two and higher.
Yet, two-dimensional holes for three-dimensional complexes are voids as the ones that may be encircled by a ball contained in $\mathcal{K}$ (i.e. cavities of the complex.) In this case, the collection of oriented two-dimensional cells on the surface of the ball represents a second homology generator of the complex.
The number of $i$-dimensional holes is often referred to as the $i$th Betti number $\beta_i(\mathcal{K})$.

In homology theory some cycles are considered the same. If two $i$-dimensional cycles (with orientation) form a boundary of a set made of $\left( i+1 \right)$-dimensional elements, then we say that they are in the same homology (equivalence) class, or simply that they are equal. Two cycles in the same homology class are represented in Fig.~\ref{fig:hom} lower left. If some cycle alone is a boundary, then we say that it is homologically trivial.

More formally, a \emph{chain} $c$ of a complex $\mathcal{K}$ is a formal sum of elements of $\mathcal{K}$ with coefficients (in this paper we consider integer or complex coefficients.) Chain $c$ is a \emph{cycle} if its boundary $\partial c$ vanishes. $c$ is \emph{homologically nontrivial} if it is not a boundary. The group of $i$-dimensional cycles is denoted by $Z_i(\mathcal{K})$. The group of $i$-dimensional boundaries is denoted by $B_i(\mathcal{K})$. The $i$th \emph{homology group} is the quotient $H_i(\mathcal{K}) = Z_i(\mathcal{K}) / B_i(\mathcal{K})$. By $H_i(\mathcal{K},\mathbb{Z})$ and $H_i(\mathcal{K},\mathbb{C})$ we denote the integer and complex homology groups, respectively.

Homology has a dual theory called cohomology. In case of complexes embedded in a tree-dimensional space, a straightforward duality between homology and cohomology generators exists. Let us fix a $i$th homology generator $c$. The dual cohomology generator $\mathbf{c}^*$ is a set of $i$th dimensional elements having the following property: every cycle in the class of $c$ needs to cross $\mathbf{c}^*$. Moreover, once the discrete analogous of integration of $\mathbf{c}^*$ on $c$ is considered by means of the dot product $\langle \mathbf{c}^*,c \rangle$ between the vectors representing the coefficients of $\mathbf{c}^*$ and $c$, the result is always $1$ (for more details consult~\cite{gstt}.) A possible cohomology generator in the considered example can be visualized in Fig.~\ref{fig:hom} upper right.
Analogously as in the case of homology, in cohomology two $\mathbf{c}^*$ and $\mathbf{c}^{*'}$ are considered equal if there exist a set $\mathbf{s}$ of $\left(i-1\right)$-dimensional elements such that $\mathbf{c}^*$ and $\mathbf{c}^{*'}$ are common \emph{co}boundary of elements of $\mathbf{s}$, see Fig.~\ref{fig:hom} bottom right for an example.

More formally, a \emph{cochain} with integer coefficients is a map from the group of chains to integers. A cochain $\mathbf{c}^*$ is a \emph{cocycle} if its coboundary $\delta \mathbf{c}^*$ vanishes. $\delta$ is the coboundary operator~\cite{massey} defined with the Generalized Stokes Theorem $\langle \delta \mathbf{c}^*,c \rangle = \langle \mathbf{c}^*, \partial c \rangle$. $\mathbf{c}^*$ is \emph{cohomologically nontrivial} if it is not a coboundary. The group of $i$-dimensional cocycles of the complex $\mathcal{K}$ is denoted by $Z^i(\mathcal{K})$, whereas the group of $i$-dimensional coboundaries is denoted as $B^i(\mathcal{K})$. The $i$th \emph{cohomology group} is the quotient $H^i(\mathcal{K}) = Z^i(\mathcal{K}) / B^i(\mathcal{K})$. By $H^i(\mathcal{K},\mathbb{Z})$ and $H^i(\mathcal{K},\mathbb{C})$ we denote integer and complex cohomology groups, respectively. By a \emph{(co)homology basis} we mean a set of (co)homology generators that span the (co)homology group.

One may also consider the so-called \textit{relative} (co)homology group of a complex $\mathcal{K}$ modulo a sub-complex $\mathcal{K}_0 \subset \mathcal{K}$ in which $\mathcal{K}_0$ is forgotten. The relative (co)homology groups are denoted by $H_i(\mathcal{K},\mathcal{K}_0)$ and $H^i(\mathcal{K},\mathcal{K}_0)$. The details can be found in~\cite{massey} or, more informally, in~\cite{gstt},~\cite{ijnme}.

The \textit{dual complex} $\tilde{\mathcal{K}}$ \cite{munkres}, \cite[Section 3]{cmame} is obtained from $\mathcal{K}$ by using the \textit{barycentric subdivision}. Let us define the dual cell complex $\tilde{\mathcal{K}}=D(\mathcal{K})$ in the following way:
\begin{enumerate}
\item For every polyhedron $t \in \mathcal{K}$, $\tilde{n}=D(t)$ is defined as the barycenter of $t$.
\item For every triangle $f \in \mathcal{K}$ that is a common face of polyhedra $t_1,t_2 \in \mathcal{K}$, $\tilde{e}=D(f)$ is defined as the sum of a segment of line joining the barycenter of $f$ with $D(t_1)$ and a segment of line joining the barycenter of $f$ with $D(t_2)$.
\item For every edge $e \in \mathcal{K}$ let $f_1,\ldots,f_n \in \mathcal{K}$ be the triangles incidental to $e$. $\tilde{f}=D(e)$ is then defined as a (set-theoretical) sum $\bigcup_{i=1}^n \mathrm{conv}[ B(e),D(f_i) ]$, where $B(e)$ denotes the barycenter of the edge $e$ and $\mathrm{conv}[\cdot]$ the convex hull.
\item For every node $n \in \mathcal{K}$ let $e_1,\ldots,e_n \in \mathcal{K}$ be the edges incidental to $n$. $\tilde{t}=D(n)$ is the volume bounded by $D(e_1),\ldots,D(e_n)$.
\end{enumerate}

\section{(Co)homology and low-frequency electrodynamics}\label{cohomologyinem}
\begin{figure}
\centering
\includegraphics[width=10cm]{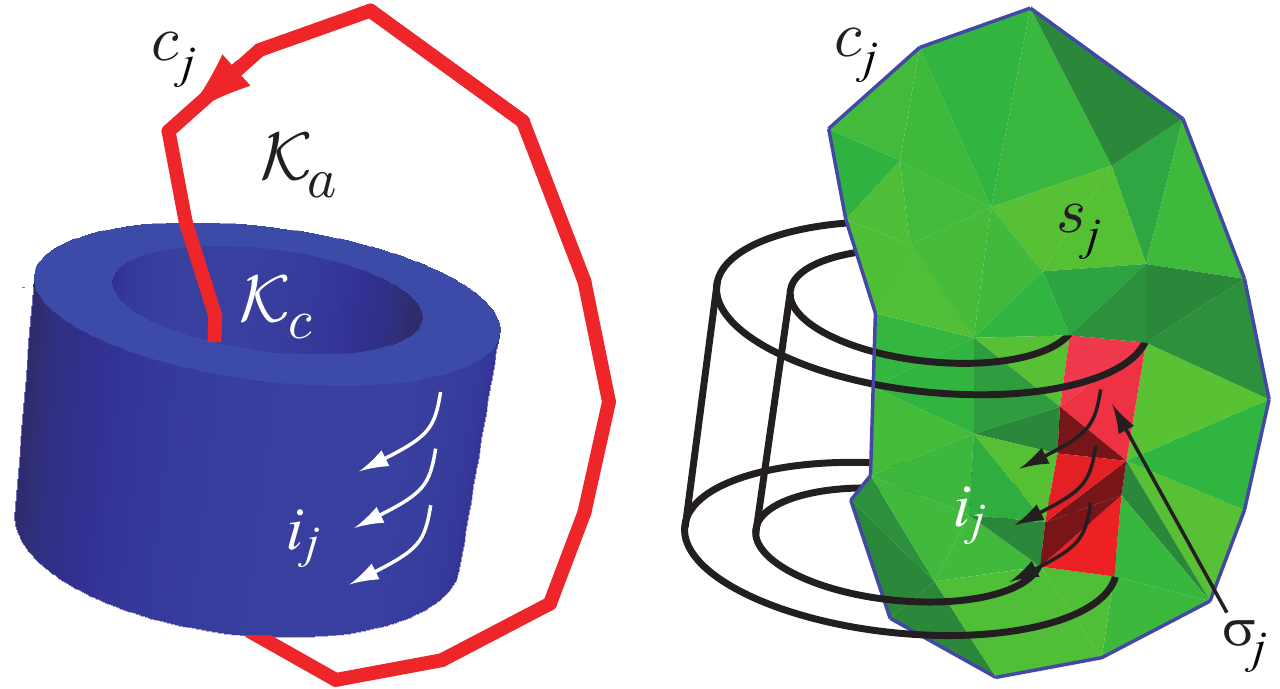}%
\caption{On the left, the independent current flowing in a torus. On the right, the $1$-cycle $c_j \in C_1(\mathcal{K}_a)$ and a $2$-chain $s_j$ such that $c_j=\partial s_j$. \label{fig1}}
\end{figure}

Let us cover the topologically trivial computational domain by a conformal polyhedral mesh $\mathcal{K}$ which is a regular CW-complex~\cite{massey}. Two sub-complexes $\mathcal{K}_c$ and $\mathcal{K}_a$ of $\mathcal{K}$ are introduced that contain mesh elements belonging to the conducting and insulating regions, respectively.
Let us define the potentials in $\mathcal{K}_a$ for a harmonic analysis of a MQS problem formulated by using a magnetic scalar potential, as the $\mathbf{T}$-$\mathbf{\Omega}$ formulation~\cite{cmame},~\cite{CiCP},~\cite{tct}. The algebraic Amp\`{e}re's law in the insulating region is enforced on every $2$-cell with $\delta \mathbf{F}=\mathbf{I}=\mathbf{0}$,
where $\mathbf{I}$ is the complex-valued electric current $2$-cochain and $\mathbf{F}$ is the magneto-motive force (m.m.f.) complex-valued $1$-cochain.
Thanks to the discrete Hodge decomposition, the $1$-cocycle $\mathbf{F} \in Z^1(\mathcal{K}_a,\mathbb{C})$ can be expressed as a linear combination of a basis of the $1$st cohomology group $H^1(\mathcal{K}_a , \mathbb{C})$ plus a $1$-coboundary $B^1(\mathcal{K}_a,\mathbb{C})$. The $1$-coboundary $B^1(\mathcal{K}_a,\mathbb{C})$ is provided by taking the $0$-coboundary of a complex-valued $0$-cochain $\boldsymbol{\Omega}$ whose coefficients represent the magnetic scalar potential sampled on mesh nodes. Since $\mathcal{K}_a$ is embedded in $\mathbb{R}^3$, the (co)homology groups are torsion free~\cite{munkres} and the basis of $H^1(\mathcal{K}_a , \mathbb{C})$ can be obtained from a basis of $H^1(\mathcal{K}_a , \mathbb{Z})$ where the elements of $\mathbb{Z}$ are treated as elements of $\mathbb{C}$~\cite{massey}. Then, the \textit{nonlocal} (i.e. applied not locally on each $2$-cell as $\delta \mathbf{F}=\mathbf{I}$, but on an arbitrary $2$-chain) algebraic Amp\`{e}re's law~\cite{cmame},~\cite{CiCP} $\langle \mathbf{F}, c_j \rangle=\langle \mathbf{I}, s_j\rangle$ (see Fig. \ref{fig1}) implicitly holds for any $1$-cycle $c_j \in C_1(\mathcal{K}_a)$, with $c_j=\partial s_j$, by setting
\begin{equation}\label{potdef1}\mathbf{F}=\delta \boldsymbol{\Omega}+\sum_{j=1}^{\beta_1(\mathcal{K}_a)} i_j\,\mathbf{h}^j,\end{equation}
where $\langle \centerdot, \centerdot\rangle$ denotes the dot product, $\{\mathbf{h}^j\}_{j=1}^{\beta_1(\mathcal{K}_a)}$ are the representatives of the $1$st cohomology group $H^1(\mathcal{K}_a, \mathbb{Z})$ generators and $\beta_1(\mathcal{K}_a)$ is the 1st Betti number of $\mathcal{K}_a$. When it is not confusing, by cohomology generators we refer to both the cohomology classes and their representatives. Physically, the $\{i_j\}_{j=1}^{\beta_1(\mathcal{K}_a)}$ can be interpreted as a set of \textit{independent currents}~\cite{cmame},~\cite{CiCP}
flowing in the branches of the conductors $\mathcal{K}_c$. Fig. \ref{fig1} shows the independent current flowing in a solid $2$-dimensional torus. For a $n$-fold $2$-dimensional solid torus, there are $n$ independent currents.

Let us fix the generators $\{\mathbf{h}^j\}_{j=1}^{\beta_1(\mathcal{K}_a)}$ of $H^1(\mathcal{K}_a, \mathbb{Z})$. There exists a set of cycles $\{c_i\}_{i=1}^{\beta_1(\mathcal{K}_a)}$ being a $H_1(\mathcal{K}_a, \mathbb{Z})$ basis such that $\langle \mathbf{h}^j , c_i \rangle = \delta_{ij}$ holds~\cite{massey},~\cite{gstt}.
Since $\mathcal{K}$ is homologically trivial, there exist $2$-chains $\{ s_i \}_{i=1}^{\beta_1(\mathcal{K}_a)} \in C_2(\mathcal{K})$ whose boundaries are the $\{ c_i \}_{i=1}^{\beta_1(\mathcal{K}_a)}$. Their restrictions $\{ \sigma_i \}_{i=1}^{\beta_1(\mathcal{K}_a)}$ to $\mathcal{K}_c$ form a $H_2( \mathcal{K}_c , \partial \mathcal{K}_c, \mathbb{Z} )$ basis~\cite{tct}, see Fig. \ref{fig1}, on the right.
The independent currents are exactly the currents linked in nonlocal algebraic Amp\`{e}re's law~\cite{cmame} by the $1$-cycles $\{ c_i \}_{i=1}^{\beta_1(\mathcal{K}_a)}$ in such a way that the $j$-th independent current can be defined with $\langle \mathbf{F}, c_j \rangle=i_j=\langle \mathbf{I}, s_j\rangle=\langle \mathbf{I}, \sigma_j\rangle$. Then, the currents linked by any other cycle in $\mathcal{K}_a$ can be obtained by linear combinations of the independent currents.

\section{The \textit{D{\l}otko--Specogna} (\texttt{DS}) algorithm}\label{sec:DSAlgorithm}
Usually, cohomology generators are found first~\cite{cmame},~\cite{CiCP} and the current distribution inside $\mathcal{K}_c$ is a result of the MQS simulation.
In principle, taking inspiration from physics, one may do it the other way around: The $j$-th $1$-cocycle $\mathbf{h}^j$ can be computed by imposing a unity current in $\sigma_j$ and a zero current in the other sigmas, where the $\{ \sigma_i \}_{i=1}^{\beta_1(\mathcal{K}_a)}$ are $H_2( \mathcal{K}_c , \partial \mathcal{K}_c, \mathbb{Z} )$ generators.
If a set of currents $\{\mathbf{t}^1,\ldots,\mathbf{t}^{\beta_1(\mathcal{K}_a)}\}$ that fulfill these constraints is constructed, finding the cohomology basis is just a matter of solving $\beta_1(\mathcal{K}_a)$ linear systems $\delta\,\mathbf{h}^j=\mathbf{t}^j$ and setting a zero coefficient of the output cochains in $\mathcal{K}_c\setminus \mathcal{K}_a$. Surprisingly, these systems can be solved in most cases by back-substitutions only, without using any integer linear solver and even a sparse matrix data structure, with the \textit{Extended Spanning Tree Technique} (\texttt{ESTT})~\cite{estt},~\cite{tct}. Intuitivelly ESTT algorithm is a way to extend a cohomology generator of a space $S \subset X$ to a cocycle in $X$.

\begin{figure}
\centering
\includegraphics[width=10cm]{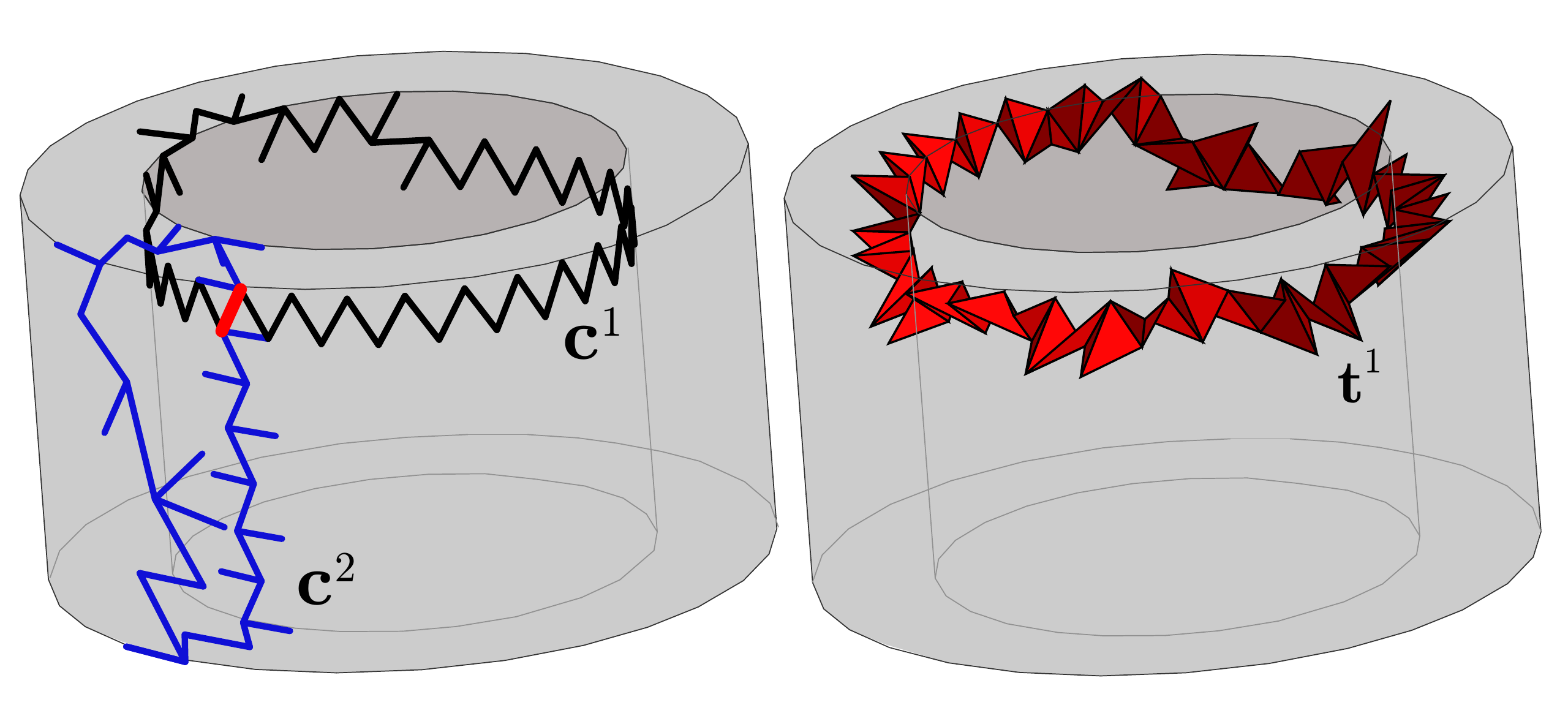}%
\caption{On the left, the two cohomology generators for a 2-torus intersecting in the thicker edge. On the right, the support of the thinned current $\mathbf{t}^1$ corresponding to $\mathbf{c}^1$. \label{fig2}}
\end{figure}
The core of this physics-based approach to cohomology computation is how to obtain such currents. The key observation is that the $1$-cocyles $\{\mathbf{h}^j\}_{j=1}^{\beta_1(\mathcal{K}_a)}$ do not depend on how the unity currents are distributed inside $\mathcal{K}_c$. Moreover, if each unity current flows in a ring whose sections are ``single $2$-cells'', then the constraint can be trivially imposed by assigning a unity current (the sign depends on incidence) to those $2$-cells. The resulting $2$-cocycles in $\mathcal{K}_c$ are called \textit{thinned currents}~\cite{tct}. An example of thinned current in a solid $2$-torus is provided on the right of Fig. \ref{fig2}.
In \texttt{TCT} algorithm~\cite{tct}, a physically-based method to construct thinned currents is employed. A thinning is applied on $\mathcal{K}_c$ in such a way that the conductors become a ``single $3$-cell thick.'' The thinned conductors can be viewed on the dual complex as a graph representing the skeleton of $\mathcal{K}_c$. By computing independent cycles on that graph, by orienting them and by considering the $2$-cells that are dual to the dual $1$-cells in the graph, the thinned currents are found. This approach presents two problems: on the one hand, this approach does not work for conductors that do not homotopically retract to a graph, on the other hand finding the skeleton is time consuming since all elements of $\mathcal{K}_c$ have to be processed.
Here is when the \texttt{DS} algorithm, summarized as follows, comes is:
\begin{enumerate}
\item Find the $n$ combinatorial $2$-manifolds that represent the connected components $\mathcal{C}_1,\ldots,\mathcal{C}_n$ of $\partial \mathcal{K}_c$ (the external boundary $\partial \mathcal{K}$ is excluded from $\mathcal{K}_c$.) This part requires $O(card(\mathcal{K}))$ time.
\item Compute the 1st cohomology $H^1(\mathcal{C}_i, \mathbb{Z})$ generators $\mathbf{c}^1,\ldots,\mathbf{c}^{2g}$ for each $i$-th connected component of $\partial \mathcal{K}_c$, where $g$ is the genus of $\mathcal{C}_i$. This can be done in linear time worst-case complexity $O(card(\partial \mathcal{K}_c)\, g)$ with the graph-theoretic algorithm presented in~\cite{h^1surface} (see App. \ref{sec:coho2Manifolds} for a short presentation.) For an example, see on the left of Fig. \ref{fig2}.
\item For every connected component $\mathcal{C}_i$, find the \textit{thinned currents} $\mathbf{t}^1,\ldots,\mathbf{t}^{2g}$ corresponding to $\mathbf{c}^1,\ldots,\mathbf{c}^{2g}$ in $O(card(\partial \mathcal{K}_c)\, g)$ with the following algorithm:\\
\begin{tabular}{l}
 \textbf{for} each $1$-cell $E$ with nonzero coefficient $c_E$ in $\mathbf{c}^i$ \\
 $\,\,\,$ \textbf{for} each $2$-cell $T \in \mathcal{K}_c$ with $E$ in the boundary\\
 $\,\,\,$ $\,\,\,$ $\langle \mathbf{t}^i, T \rangle += c_E \kappa(T,E)$;
\end{tabular}\\
    The value of the cochain $\mathbf{t}$ on a cell $E$ is $\langle \mathbf{t},E \rangle$, whereas $\kappa(A,B)$ denotes the incidence between cells $A$ and $B$, see~\cite{massey}. Initially, set $\langle \mathbf{t}^i, T \rangle = 0$ for all $2$-cells $T \in \mathcal{K}_c$. For an example, see on the right of Fig. \ref{fig2}. The demonstration of correctness of this approach is presented in App. \ref{app2}.
\item For every connected component $\mathcal{C}_i$, solve the integer systems $\delta\, \mathbf{h}^j =\mathbf{t}^j$, $j \in \{1,\ldots,2g\}$, to find the $2g$ $1$-cocycles $\mathbf{h}^1,\ldots,\mathbf{h}^{2g}$ in $\mathcal{K}_a$. This can be performed without solving any system by a simultaneous application of the \texttt{ESTT} algorithm~\cite{estt},~\cite{tct}. Simultaneous means that the \texttt{ESTT} algorithm is applied to all $\mathbf{t}^1,\ldots,\mathbf{t}^{2g}$ thinned currents at the same time. Algorithmically this can be easily achieved by changing a real number to a vector of $2g$ real numbers in the \texttt{ESTT} algorithm.
\item For every connected component $\mathcal{C}_i$, store the restrictions of $\mathbf{h}^1,\ldots,\mathbf{h}^{2g}$ to $\mathcal{K}_a$. The average computational effort required is $O(card(\mathcal{K})\,g)$.
\end{enumerate}
If the genus $g$ is bounded by a constant $O(1)$, as it happens always in practical problems, the average complexity of the \texttt{DS} algorithm is linear $O(card(\mathcal{K}))$. It is also purely graph-theoretic, so straightforward to implement.

It is easy to prove that the $1$-cocycles obtained by the \texttt{DS} algorithm span $H^1(\mathcal{K}_a, \mathbb{Z})$. However, the obtained cocycles are not a basis of $H^1(\mathcal{K}_a, \mathbb{Z})$, since the number of obtained generators is twice the cardinality of its own basis. In other words, some cocycle is a linear combination of the others. For this reason we refer to these redundant cocycles produced by the \texttt{DS} algorithm as \emph{lazy cohomology generators}.

To obtain a $H^1(\mathcal{K}_a, \mathbb{Z})$ cohomology basis with the \texttt{DS} algorithm, only the cohomology generators on $\partial \mathcal{K}_c$ that are extended with the \texttt{ESTT} algorithm to coboundaries in $\mathcal{K}_a$ have to be used. They may be found by a change of basis obtained by computing linking numbers and a SNF of a matrix (see App. \ref{app3} for details.) Due to conceptual and implementation difficulties, one welcomes the possibility to bypass this additional step.

Another contribution of this paper is to show that in computational physics avoid partitioning the $H^1(\mathcal{C}_i, \mathbb{Z})$ generators is not only possible but even more convenient.
This is realized by employing directly the lazy cohomology generators in the physical modeling, something that has been never considered in the literature. 
Lazy cohomology generators are employed in a MQS formulation, for example the $\mathbf{T}$-$\mathbf{\Omega}$~\cite{sst},~\cite{cmame},~\cite{CiCP}, as if they were a set of standard $H^1(\mathcal{K}_a, \mathbb{Z})$ generators.
Namely, a nonlocal Faraday's equation~\cite{cmame},~\cite{CiCP} is written on the support of the $j$-th cohomology generator as $\langle \tilde{\mathbf{U}},\partial \tilde{h}_j \rangle=-i\,\omega\,\langle \tilde{\mathbf{\Phi}},\tilde{h}_j \rangle$, where $\tilde{\mathbf{U}}$ is the electro-motive force $1$-cochain on the dual complex, $\tilde{\mathbf{\Phi}}$ is the magnetic flux $2$-cochain on the dual complex and $\tilde{h}_j=D(\mathbf{h}^j)$, $D$ being the \textit{dual map}~\cite{munkres} that maps elements of the original complex to elements of the dual complex.
Lazy cohomology generators contain a $H^1(\mathcal{K}_a \mathbb{Z})$ basis and generators that are dependent to the basis. Therefore, adding the dependent equations is not a problem considering that the system of equations to solve is already overdetermined. This is due to the fact that algebraic Faraday's equations~\cite{cmame},~\cite{CiCP} enforced in $\mathcal{K}_c$ are also dependent. Even though a full-rank system may be obtained by a tree-cotree gauging (i.e. set the electric vector potential on a tree of $1$-cells in $\mathcal{K}_c$ to zero), it is widely known that with iterative linear solvers it is much more efficient to use an \textit{ungauged}~\cite{ungauged} formulation. What is important from the modeling point of view is that even if the potentials are not unique, the fields are. Therefore, the use of linearly dependent cocycles in the physical modeling does not introduce either any inconsistency in the formulation of the boundary value problem or any penalties in the computational time employed by the simulation due, for example, to a hypothetical increase of the condition number of the linear system matrix or to the use of twice as many cohomology generators as needed.

As a final observation about lazy generators, let us now consider a lazy generator belonging to the trivial class of $H^1(\mathcal{K}_a, \mathbb{Z})$. Given an arbitrary $1$-cycle $c \in Z_1(\mathcal{K}_a)$, the dot product of the lazy generator with $c$ is zero. Therefore, trivial generators verify trivially the nonlocal algebraic Amp\`{e}re's law and, in this case, the current $i_j$ does not represent the current linked by the dual homology generator.
Therefore, the value of the independent current relative to a trivial generator is not unique and it is determined by the solution of the system of equations. This is not surprising, since the independent current in this case does not have a physical meaning.

\section{Numerical experiments}\label{numres}
The \texttt{DS} algorithm has been integrated into the research software CDICE \cite{cdice} implemented in Fortran 90. Three competing algorithms presented in~\cite{cmame},~\cite{cmes} and~\cite{tct} are considered. Two computers are used to the computations: Intel Core 2 Duo T7700 2.4GHz laptop with 4GB of RAM and $64$ GB of RAM and Intel Xenon E7-8830 2.13 GHz processors computer. As the mesh size increases, standard cohomology computations end up in failures due to having exceeded the memory limit of the laptop. When the laptop runs out of memory, the large computer is used. The \texttt{DS} algorithm has been executed on the laptop up to five millions of tetrahedra without encountering any problem, which shows that it is quite economical in terms of memory usage.

Table~\ref{tab1} reports the time required (in seconds) for cohomology computation by the various algorithms. The timings obtained on the large computer are given in brackets.
The Table presents also the time that \texttt{DS} algorithm requires for computing a standard cohomology basis (see App. \ref{app3} for implementation details.)
\begin{table*}
\caption{Time required (in seconds) for cohomology computation with various algorithms. \label{tab1}}
{\scriptsize
\begin{tabular}{|l|l|l|l|l|l|l|l|l|}
\hline
Benchmark & $\beta_1(\mathcal{K}_a)$ & tetrahedra & $H^1(\mathcal{K}_a,\mathbb{Z})$ & $H^1(\mathcal{K}_a,\mathbb{Z})$ & \texttt{TCT} & \texttt{DS} & \texttt{DS}\\
 &  &  & \cite{cmame},\cite{gstt} &  \cite{cmes} & \cite{tct} & lazy & $H^1(\mathcal{K}_a,\mathbb{Z})$\\
\hline
\hline
trefoil knot & 1 & $199,208$ & $24.38$ & $23$ & $0.6$& $0.3$& $1.1$\\
\hline
spiral & 1 & $1,842,070$ & $(424.14)$ & $(612)$ & $10.1$& $1.7$& $4.1$\\
\hline
micro-inductor & 1 & $2,197,192$ & $(59359)$ & $(>70000)$ & $24.5$& $2.4$& $4.2$\\
\hline
micro-transformer & 2 & $2,582,830$ & $(>70000)$ & $(>70000)$ & $32.8$& $3.6$& $7.6$\\
\hline
micro-coaxial line & 6 & $4,861,655$ & $(612828)$ & $(6128)$ & $86.1$& $10.6$& $26.8$\\
\hline
toroidal shell & 2 & $2,769,200$ & $(1503)$ & $(>70000)$ & $(>70000)$ & $3.4$& $3.9$\\
\hline
\end{tabular}}
\end{table*}

\begin{figure}
\centering
\includegraphics[width=14cm]{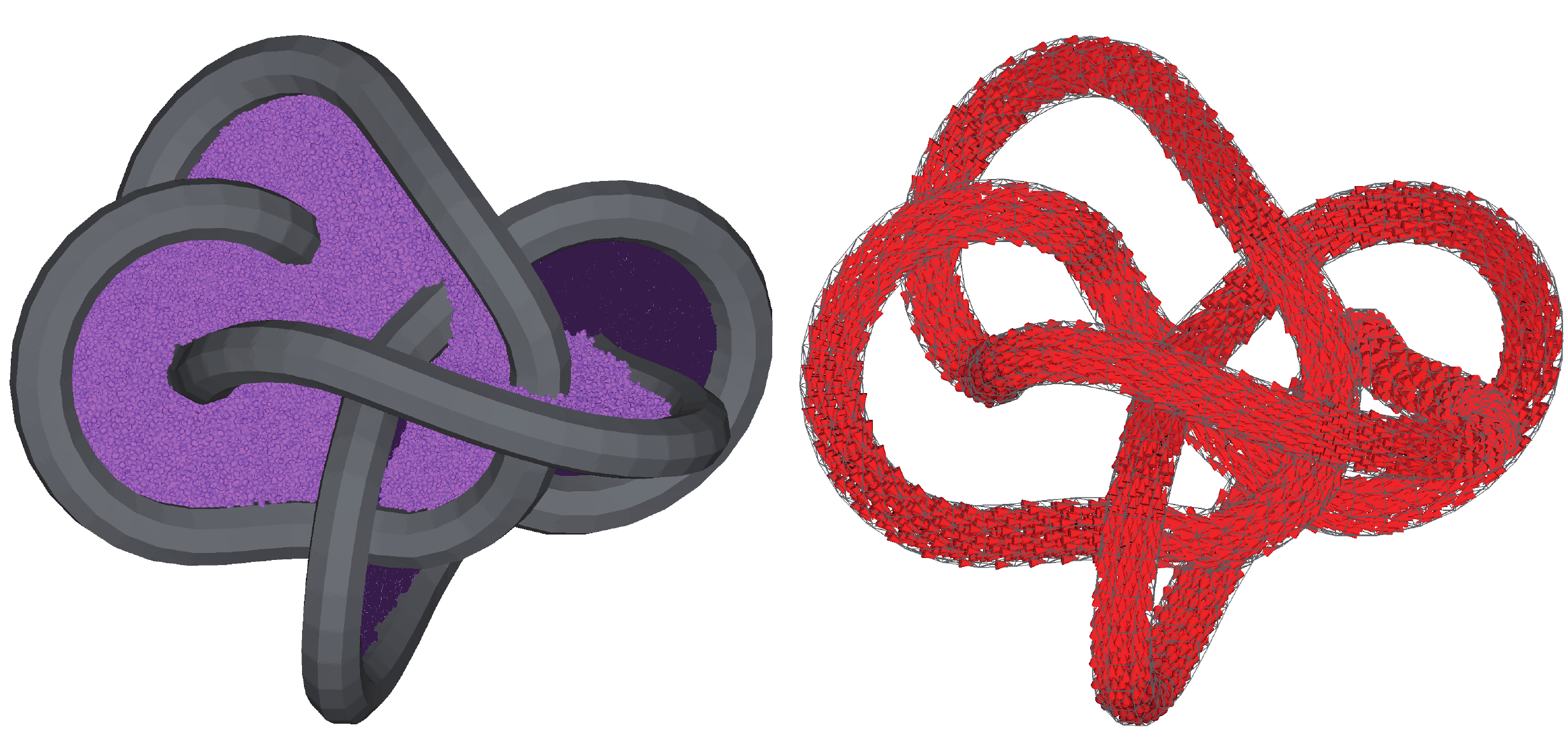}%
\caption{On the left, a complicated thick knot and the first cohomology group generator of its complement. For clarity, the dual faces in the support of $\tilde{h}_j=D(\mathbf{h}^j)$ are shown. On the right, the current density obtained by the MQS solver. \label{fig3}}
\end{figure}

The algorithm is going to be exploited to solve MQS problems arising in fusion engineering and design, in engineering and optimization of electromagnetic devices and the analysis of features of magnetic fields generated by current-carrying thick knots, see Fig. \ref{fig3}.

We remark that the \texttt{DS} algorithm may be used also in applications outside computational physics where only the complex $\mathcal{K}_a$ is available. In that case, in fact, one may employ an efficient mesh generator as TetGen (\verb"http://www.tetgen.org/") to produce the mesh (without adding Steiner points~\cite{bern}) of the complement with respect to a box containing $\mathcal{K}_a$.

\section{Conclusions}\label{conclusions}
The \texttt{DS} algorithm, even though is general and straightforward to implement, outperforms all competing state-of-the-art algorithms for first cohomology group computations.
The time required for computing cohomology generators with the \texttt{DS} algorithm is so limited that it allows to remove what has been considered for more than twenty-five years the main simulation bottleneck for low-frequency electrodynamics problems. Therefore, we expect the \texttt{DS} algorithm to be embedded in the next-generation electromagnetic CAE softwares.

\appendix

\section{Cohomology generators of $2$-manifolds}
\label{sec:coho2Manifolds}
In this section the algorithm to compute $H^1(\partial \mathcal{K}_c)$ generators presented in~\cite{h^1surface} is recalled. For simplicity, we consider 2-manifolds without boundary only. The extension to the general case can be found in~\cite{h^1surface}.

By the \emph{primal skeleton} of $\partial \mathcal{K}_c$ we mean the graph consisting of all the vertices and edges in $\partial \mathcal{K}_c$. By the \emph{dual skeleton} of $\partial \mathcal{K}_c$ we mean a graph whose vertices are the 2-cells in $\partial \mathcal{K}_c$ and an edge is put between two vertices iff. the corresponding faces in $\partial \mathcal{K}_c$ share an edge in $\partial \mathcal{K}_c$. We want to point out that edges of both the primal and dual skeletons correspond to edges of $\partial \mathcal{K}_c$.

Let us fix a spanning tree $T$ of the primal skeleton. Let us also fix a spanning tree $T'$ of dual skeleton. We assume, that $T$ and $T'$ do not share edges of $\partial \mathcal{K}_c$. In~\cite{h^1surface} and~\cite{erickson} it is shown that the number of edges in $\partial \mathcal{K}_c$ that are neither in $T$ not in $T'$ is the first Betti number of $\partial \mathcal{K}_c$. Moreover, the $H_1(\partial \mathcal{K}_c)$ generators are the cycles closed in $T$ by those edges, whereas the $H^1(\partial \mathcal{K}_c)$ generators are the cycles closed in $T'$ by those edges.

\begin{figure}[!h]
\centering
\includegraphics[width=10cm]{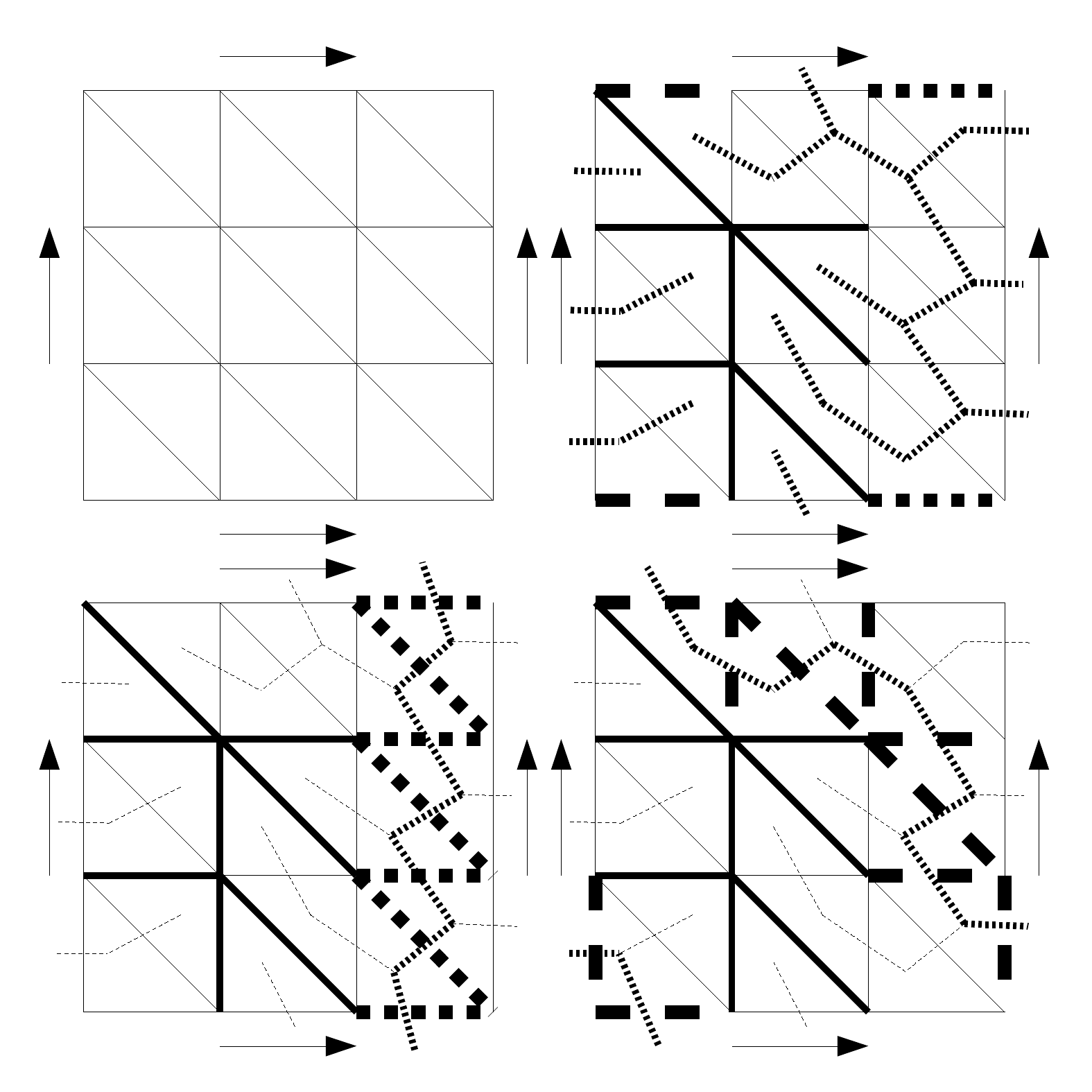}
\caption{Upper left, the standard triangulation of a torus. Opposite sides are identified. Upper right, tree on primal (solid bold), and dual (dotted bold) skeleton. With double bold, the two edges not belonging to one of the trees are depicted. Lower left, a cohomology generator closed by the first edge. Lower right, cohomology generator closed by the second edge.}
\label{fig:cohoOnBd}
\end{figure}

The idea of the procedure is presented in Figure~\ref{fig:cohoOnBd}.

In order to obtain the coefficients of the cocycle, a simple procedure that orients the cycle is used. Let $v$ and $w$ be the vertices of the edge that close the cycle. With a BFS strategy~\cite{cormen} a distance function on the tree (or dual tree) from $v$ is built as long as $w$ is not reached. Then, a path in a tree from $w$ to $v$ is found by following the decreasing values of the defined function. The obvious details are left for the reader.

\section{Thinned currents construction}\label{app2}
In this section we show that the output $t$ of the algorithm to obtain thinned currents from $H^1(\partial \mathcal{K}_c)$ is indeed a cochain (this is a necessary and sufficient condition for the ESTT algorithm termination~\cite{estt}.) To do this, first we need to remind one of Massey's equation~\cite{massey} which is a property of any regular CW-complex. Let $T$ be a 3-dimensional cell having 1-dimensional cell $E$ in boundary. Then there exist exactly two 2-dimensional cells $T_1$ and $T_2$ in boundary of $T$ both having $E$ in their boundary. Moreover, the incidence indices satisfy the following equation (Theorem IX.7.2 in~\cite{massey}):
\[ \kappa(K,T_1) \kappa(T_1,E) + \kappa(K,T_2) \kappa(T_2,E) = 0.\]
In order to show that $t$ is a cocycle, we have to show that $\delta t = 0$. This is equivalent to showing that for every 3-dimensional cell $T$, $\langle \delta t , T \rangle = 0$. Of course, due to the the algorithm to obtain thinned currents from $H^1(\partial \mathcal{K}_c)$, this expression may be nonzero only for 3-dimensional cells in $\mathcal{K}_c$ that have some edge(s) from in input 1-cocycle $c$ in boundary. To show that in this case $\langle \delta t , T \rangle = 0$, we need to consider the following cases:
\begin{enumerate}
\item $T$ has only one edge $E$ in the boundary such that $\langle c , E \rangle \neq 0$. Then the two 2-cells $T_1$ and $T_2$ from the Massey's equation will be nonzero in the cochain $t$ (we have $\langle t , T_1 \rangle = \langle c, E \rangle \kappa(T_1,E)$ and $\langle t, T_2 \rangle = \langle c, E \rangle \kappa(T_2,E)$ respectively.) From the Massey's equation, we know that $\kappa(T,T_1) \kappa(T_1,E) + \kappa(T,T_2) \kappa(T_2,E) = 0$. After multiplying this equation by $\langle c, E \rangle$ we get
\[\kappa(T,T_1)\kappa(T_1,E)\langle c, E \rangle + \kappa(T,T_2)\kappa(T_2,E)\langle c, E \rangle=0.\]
Therefore,
\begin{eqnarray*}
& 0 = \kappa(T,T_1) (\kappa(T_1,E)\langle c, E \rangle) + \\
& \kappa(T,T_2) (\kappa(T_2,E) \langle c, E \rangle) = \kappa(T,T_1) \langle t, T_1 \rangle +\\
& \kappa(T,T_2) \langle t, T_2 \rangle = \langle \delta t , T \rangle
\end{eqnarray*}
that prove this case.
\item $T$ has two edges $E_1$ and $E_2$ in the boundary such that $\langle c, E_1 \rangle \neq 0$ and $\langle c, E_2 \rangle \neq 0$. We assume that $E_1$ and $E_2$ are two constitutive edges in cocycle $c$, therefore they need to share a 2-cell $T_3$ in $\partial \mathcal{K}_c$. Since $c$ is a cocycle, we have
\begin{equation}
\label{eq:cochain}
\langle E_1,c \rangle \kappa(T_3,E_1) + \langle E_2 , c \rangle \kappa(T_3,E_2) = 0.
\end{equation}
Moreover, we consider the following Massey's equations
\begin{equation}
\label{eq:mass1}
\kappa(T,T_1)\kappa(T_1,E_1) + \kappa(T,T_3)\kappa(T_3,E_1) = 0,
\end{equation}
\begin{equation}
\label{eq:mass2}
\kappa(T,T_2)\kappa(T_2,E_2) + \kappa(T,T_3)\kappa(T_3,E_2) = 0.
\end{equation}
From~(\ref{eq:cochain}), we have
\begin{eqnarray*}
0 = -\kappa(T,T_3)(\langle E_1,c \rangle \kappa(T_3,E_1) + \langle E_2 , c \rangle \kappa(T_3,E_2)) = \\
(-\kappa(T,T_3) \kappa(T_3,E_1)) \langle E_1,c \rangle  + (-\kappa(T,T_3) \kappa(T_3,E_2)) \langle E_2 , c \rangle.
\end{eqnarray*}
From~(\ref{eq:mass1}) and~(\ref{eq:mass2}), we have
\begin{eqnarray*}
(-\kappa(T,T_3) \kappa(T_3,E_1)) \langle E_1,c \rangle + (-\kappa(T,T_3) \kappa(T_3,E_2)) \langle E_2 , c \rangle =\\
\kappa(T,T_1) (\kappa(T_1,E_1) \langle E_1,c \rangle) + \kappa(T,T_2) (\kappa(T_2,E_2)\langle E_2 , c \rangle) = \\
\kappa(T,T_1) \langle t , T_1 \rangle + \kappa(T,T_2) \langle t , T_2 \rangle = \langle \delta t, T \rangle,
\end{eqnarray*}
that prove this case.
\item If $T$ has more than two edges in the boundary that have nonzero value in cochain $c$, then the conclusion follows from the inductive application of the argument for two edges. The obvious details are left for the reader.
\end{enumerate}
\hspace{1cm}

\section{Finding cocycles in $\partial \mathcal{K}_c$ that extend to a cohomology basis of $\mathcal{K}_a$}\label{app3}
A technique to partition the homology generators on a combinatorial $2$-manifold into two classes---the ones that bound in $\mathcal{K}_c$, and the ones that bound in $\mathcal{K}_a$---is proposed in~\cite{hiptmair2d}. This algorithm is based on the computation of linking numbers between all possible pairs of generators and subsequent SNF computation of a small integer matrix containing the computed linking numbers.
Linking numbers are defined for disjoint $1$-cycles only, so a pre-processing has been applied in~\cite{hiptmair2d} on surface generators to perturb them in such a way no intersections are present between any pair of generators. For this purpose, the \textit{submerged cycles} and \textit{shifted surface cycles} have been found. This is not completely trivial to do algorithmically and cycles have to be ``straightened'' before~\cite{hiptmair2d}. The complexity of the linking number computation is quadratic with the number of edges in $\partial \mathcal{K}_c$ and a computationally costly interval arithmetic~\cite{hayes} package has to be used to rigorously compute linking numbers without the risk of errors due to the finite precision of real numbers~\cite{hayes},~\cite{aral}, especially on coarse meshes.

In this paper, to find the change of basis to find cohomology generators in $\mathcal{K}_a$ we use the same idea presented in~\cite{hiptmair2d}, but finding linking numbers between paths on the dual complex~\cite{munkres},~\cite{cmame}. This approach, contrarily to~\cite{hiptmair2d}, does not require any extra computations since the surface cycles are obtained by considering the dual $1$-cells that are dual to $1$-cells in the support of the cohomology generators, while the submerged cycles are defined simply as the dual $1$-cells that are dual to $2$-cells in the support of thinned currents. We want to remark that for this purpose the $2$-cells belonging to the support of thinned currents have to be in order. They can be easily ordered in linear time and the technical details are left for the reader.

About practical implementation of the change of basis, the algorithm presented in~\cite{aral} is used to compute linking numbers and the interval arithmetic library for Fortran 90 presented in~\cite{interval} is employed for interval arithmetic computations.


\bibliography{cpc_ds}{}

\begin{thebibliography}{10}

\bibitem{munkres}
JR. Munkres.
\newblock {\em Elements of algebraic topology}.
\newblock Perseus Books, Cambridge, MA, 1984.

\bibitem{dey1}
T.H. Dey, K.~Li, J.~Sun, and D.~Cohen-Steiner.
\newblock Computing geometry-aware handle and tunnel loops in 3d models.
\newblock {\em ACM Trans. Graph.}, 27:1--9, 2008.

\bibitem{gu3}
X.~Guo, X.~Li, Y.~Bao, X.~Gu, and H.~Qin.
\newblock Meshless thin-shell simulation based on global conformal
  parameterization.
\newblock {\em IEEE Trans. Vis. Comput. Gr.}, 12:375--385, 2006.

\bibitem{pattern}
M.~Mrozek, M.~\.{Z}elawski, A.~Gryglewski, S.~Han, and A.~Krajniak.
\newblock Homological methods for extraction and analysis of linear features in
  multidimensional images.
\newblock {\em Pattern Recogn.}, 45(1):285--298, January 2012.

\bibitem{sensors}
Vin De~Silva and R.~Ghrist.
\newblock Homological sensor networks.
\newblock {\em Notices of the AMS}, 54, 2007.

\bibitem{robotmotionplanning}
M.~Farber.
\newblock Topological complexity of motion planning.
\newblock {\em Discr. \& Comput. Geom.}, 29(2):211--221, 2003.

\bibitem{cancer}
Monica Nicolau, Arnold~J. Levine, and Gunnar Carlsson.
\newblock {Topology based data analysis identifies a subgroup of breast cancers
  with a unique mutational profile and excellent survival}.
\newblock {\em Proceedings of the National Academy of Sciences},
  108(17):7265--7270, April 2011.

\bibitem{mezey}
P.G. Mezey.
\newblock Group theory of electrostatic potentials: A tool for quantum chemical
  drug design.
\newblock {\em Int. J. Quant. Chem.}, 28:113--122, 1985.

\bibitem{k1}
K.~Mischaikow, M.~Mrozek, J.~Reiss, and A.~Szymczak.
\newblock Construction of symbolic dynamics from experimental time series.
\newblock {\em Phys. Rev. Lett.}, 82:1144--1147, Feb 1999.

\bibitem{k2}
Marcio Gameiro, Konstantin Mischaikow, and William Kalies.
\newblock Topological characterization of spatial-temporal chaos.
\newblock {\em Phys. Rev. E}, 70:035203, Sep 2004.

\bibitem{k3}
H\"useyin Kurtuldu, Konstantin Mischaikow, and Michael~F. Schatz.
\newblock Extensive scaling from computational homology and karhunen-lo\`eve
  decomposition analysis of rayleigh-b\'enard convection experiments.
\newblock {\em Phys. Rev. Lett.}, 107:034503, Jul 2011.

\bibitem{distillation}
H.~Bombin and M.~A. Martin-Delgado.
\newblock Topological quantum distillation.
\newblock {\em Phys. Rev. Lett.}, 97:180501, Oct 2006.

\bibitem{jurke}
B.~Jurke.
\newblock Computing cohomology on toric varieties.
\newblock {\em arXiv:1109.1571 [math.AG]}, 2012.

\bibitem{kotiuga}
P.W. Gross and P.~R. Kotiuga.
\newblock {\em Electromagnetic Theory and Computation: A Topological Approach}.
\newblock MSRI Vol. 48, Cambridge University Press, 2004.

\bibitem{sst}
R.~Specogna, S.~Suuriniemi, and F.~Trevisan.
\newblock Geometric $t$-$\omega$ approach to solve eddy-currents coupled to
  electric circuits.
\newblock {\em Int. J. Numer. Meth. Eng.}, 74:101--115, 2008.

\bibitem{cmame}
P.~D{\l}otko, R.~Specogna, and F.~Trevisan.
\newblock Automatic generation of cuts on large-sized meshes for $t$-$\omega$
  geometric eddy-current formulation.
\newblock {\em Comput. Meth. Appl. Mech. Eng.}, 198:3765--3781, 2009.

\bibitem{cmes}
P.~D{\l}otko and R.~Specogna.
\newblock Efficient cohomology computation for electromagnetic modeling.
\newblock {\em CMES}, 60:247--278, 2010.

\bibitem{ijnme}
R.~Specogna.
\newblock Complementary geometric formulations for electrostatics.
\newblock {\em Int. J. Numer. Meth. Eng.}, 86:1041--1068, 2011.

\bibitem{CiCP}
P.~D{\l}otko and R.~Specogna.
\newblock Cohomology in 3d magneto-quasistatic modeling.
\newblock {\em Commun. Comput. Phys. (arXiv:1111.2374)}, 2012.

\bibitem{bestSmith}
A.~Storjohann.
\newblock Near optimal algorithms for computing smith normal form of integer
  matrices.
\newblock In {\em Proceedings of the 1996 international symposium on symbolic
  and algebraic computation ISAAC}, pages 267--274, 1996.

\bibitem{tct}
P.~D{\l}otko and R.~Specogna.
\newblock {\em Comput. Meth. Appl. Mech. Eng.}, pages
  http://dx.doi.org/10.1016/j.cma.2012.08.009, in press.

\bibitem{bott}
R.~Bott and L.W. Tu.
\newblock {\em Differential Forms in Algebraic Topology}.
\newblock Springer-Verlag, 1982.

\bibitem{maxwell}
J.~C. Maxwell.
\newblock {\em A Treatise on Electricity and Magnetism}.
\newblock Clarendon Press, Oxford, 1891.

\bibitem{kotiugaap}
P.~R. Kotiuga.
\newblock On making cuts for magnetic scalar potentials in multiply connected
  regions.
\newblock {\em J. Appl. Phys.}, 61:3916--3918, 1987.

\bibitem{massey}
W.S. Massey.
\newblock {\em A Basic Course in Algebraic Topology, vol. 127 of Graduate Texts
  in Mathematics}.
\newblock Springer-Verlag, 1991.

\bibitem{gstt}
P.~D{\l}otko and R.~Specogna.
\newblock Critical analysis of the spanning tree techniques.
\newblock {\em SIAM J. Numer. Anal.}, 48:1601--1624, 2010.

\bibitem{estt}
P.~D{\l}otko and R.~Specogna.
\newblock Efficient generalized source field computation for $h$-oriented
  magnetostatic formulations.
\newblock {\em Eur. Phys. J.-Appl. Phys.}, 53:20801, 2011.

\bibitem{h^1surface}
P.~D{\l}otko.
\newblock A fast algorithm to compute cohomology group generators of orientable
  $2$-manifolds.
\newblock {\em Pattern Recogn. Lett.}, 33:1468--1476, 2012.

\bibitem{ungauged}
Z.~Ren and A.~Razek.
\newblock Comparison of some 3d eddy current formulations in dual systems.
\newblock {\em IEEE Trans. Magn.}, 36:751--755, 2000.

\bibitem{cdice}
R.~Specogna.
\newblock Cdice research software, http://www.comphys.com, 2008--2012.

\bibitem{bern}
Marshall Bern.
\newblock Compatible tetrahedralizations.
\newblock In {\em Proceedings of the ninth annual symposium on Computational
  geometry}, SCG '93, pages 281--288, New York, NY, USA, 1993. ACM.

\bibitem{erickson}
J.~Erickson and K.~Whittlesey.
\newblock Greedy optimal homotopy and homology generators.
\newblock {\em Proc. 16th Annual ACM–SIAM Symposium Discrete Algorithms
  (SODA)}, pages 1038--1046, 2005.

\bibitem{cormen}
Thomas~H. Cormen, Charles~E. Leiserson, Ronald~L. Rivest, and Clifford Stein.
\newblock {\em Introduction to Algorithms}.
\newblock The MIT Press, 2 edition, 2001.

\bibitem{hiptmair2d}
R.~Hiptmair and J.~Ostrowski.
\newblock Generators of $h_1(\gamma_h,\mathbb{Z})$ for triangulated surfaces:
  construction and classification.
\newblock {\em SIAM J. Comput.}, 31(5):1405--1423.

\bibitem{hayes}
Brian Hayes.
\newblock A lucid interval.
\newblock {\em American Scientist}, 91(6):484--488, November--December 2003.

\bibitem{aral}
Zin Arai.
\newblock A rigorous numerical algorithm for computing the linking number of
  links.
\newblock {\em Nonlinear Theory and Its Applications}, in press.

\bibitem{interval}
R.~Baker Kearfott.
\newblock {Algorithm 763}: {INTERVAL\_ARITHMETIC}: {A Fortran 90} module for an
  interval data type.
\newblock {\em {ACM} Transactions on Mathematical Software}, 22(4):385--392,
  December 1996.

\end{thebibliography}
\bibliographystyle{unsrt}

\end{document}